\documentclass[10pt,letter,prl,twocolumn,nofootinbib,hypertext]{revtex4-1}
\usepackage{ifthen}
\newboolean{uprightparticles}
\setboolean{uprightparticles}{false} %Set to true to get roman particle symbols

\usepackage{amssymb}
\usepackage{amsfonts}
\usepackage{upgreek}
\usepackage{xspace}
\usepackage{color}
\usepackage{amsmath}
\usepackage{graphicx}
\usepackage{bm}
\usepackage{ctable}
\usepackage{placeins}

\newboolean{articletitles}
\setboolean{articletitles}{true}

\newboolean{pdflatex}
\setboolean{pdflatex}{true} % use this if using non-eps figures

\newcommand{\BRof}[1]{\ensuremath{{\cal B}(#1)}\xspace}
\newcommand{\Bsmumu}{\ensuremath{B^0_s \to\mu^+\mu^-}\xspace}
\newcommand{\Bdmumu}{\ensuremath{B^0\to\mu^+\mu^-}\xspace}
\newcommand{\Bmumu}{\ensuremath{B^{0}_{(s)}\to\mu^+\mu^-}\xspace}
\newcommand{\Bsmm}{\ensuremath{B^0_s\to\mu^+\mu^-}\xspace}
\newcommand{\Bdmm}{\ensuremath{B^0\to\mu^+\mu^-}\xspace}

\def\B       {\ensuremath{B}\xspace}
\def\Bd      {\ensuremath{B^0}\xspace}
\def\Bs      {\ensuremath{B^0_s}\xspace}
\newcommand{\mBd}{\ensuremath{m_{\Bd}}\xspace}
\newcommand{\mBs}{\ensuremath{m_{\Bs}}\xspace}

\def\lhcb {LHCb\xspace}
\newcommand{\Bhh}{\ensuremath{B^0_{(s)}\to h^+{h'}^{-}}\xspace}
\newcommand{\IP}{\ensuremath{{\rm IP}}\xspace}

\newcommand{\BuJpsiK}{\ensuremath{B^+\to J/\psi K^+}\xspace}
\newcommand{\BsJpsiPhi}{\ensuremath{B^0_s\to J/\psi \phi}\xspace}
\newcommand{\BdKpi}{\ensuremath{B^0\to K^+\pi^-}\xspace}
\newcommand{\BsKK}{\ensuremath{B^0_s\to K^+K^-}\xspace}
\newcommand{\bbdim}{\ensuremath{b\bar{b}\to \mu^+ \mu^- X}\xspace}
\newcommand{\CLsb}{\ensuremath{\textrm{CL}_{\textrm{s+b}}}\xspace}
\newcommand{\CLs}{\ensuremath{\textrm{CL}_{\textrm{s}}}\xspace}
\newcommand{\CLb}{\ensuremath{\textrm{CL}_{\textrm{b}}}\xspace}

\newcommand{\gevc}{\ensuremath{{\mathrm{\,Ge\kern -0.1em V\!/}c}}\xspace}
\newcommand{\mevc}{\ensuremath{{\mathrm{\,Me\kern -0.1em V\!/}c}}\xspace}
\newcommand{\gevcc}{\ensuremath{{\mathrm{\,Ge\kern -0.1em V\!/}c^2}}\xspace}
\newcommand{\gevgevcccc}{\ensuremath{{\mathrm{\,Ge\kern -0.1em V^2\!/}c^4}}\xspace}
\newcommand{\mevcc}{\ensuremath{{\mathrm{\,Me\kern -0.1em V\!/}c^2}}\xspace}

\def\Y#1S{\ensuremath{\Upsilon{(#1S)}}\xspace}% no space before {...}!

\newcommand\TTstrut{\rule{0pt}{3.2ex}}

\newcommand\BBstrut{\rule[-1.8ex]{0pt}{0pt}}

\def\BF         {{\ensuremath{\cal B}\xspace}}

\def\invfb   {\ensuremath{\mbox{\,fb}^{-1}}\xspace}
\newcommand{\tev}{\ensuremath{\mathrm{\,Te\kern -0.1em V}}\xspace}
\newcommand{\CL}{CL\xspace}
\usepackage{hyperref}
\usepackage[all]{hypcap}

\usepackage{mciteplus}
\begin{document}
\title{Strong constraints on the rare decays \mbox{\boldmath $B^0_s \rightarrow \mu^+\mu^-$}
and \mbox{\boldmath $B^0 \rightarrow \mu^+\mu^-$}\\ \vspace{3mm}}
\author{%%%%%%%%%%%%%%%%%%%%%%%%%%%%%%%%%%%%%%%%%%
\centerline{\large\bf The LHCb collaboration}
\begin{flushleft}
\small
R.~Aaij$^{38}$, 
C.~Abellan~Beteta$^{33,n}$, 
A.~Adametz$^{11}$, 
B.~Adeva$^{34}$, 
M.~Adinolfi$^{43}$, 
C.~Adrover$^{6}$, 
A.~Affolder$^{49}$, 
Z.~Ajaltouni$^{5}$, 
J.~Albrecht$^{35}$, 
F.~Alessio$^{35}$
% ...

{\footnotesize \it
% ...
$ ^{5}$Clermont Universit\'{e}, Universit\'{e} Blaise Pascal, CNRS/IN2P3, LPC, Clermont-Ferrand, France\\
$ ^{6}$CPPM, Aix-Marseille Universit\'{e}, CNRS/IN2P3, Marseille, France\\
% ...
$ ^{11}$Physikalisches Institut, Ruprecht-Karls-Universit\"{a}t Heidelberg, Heidelberg, Germany\\
% ...
$ ^{33}$Universitat de Barcelona, Barcelona, Spain\\
$ ^{34}$Universidad de Santiago de Compostela, Santiago de Compostela, Spain\\
$ ^{35}$European Organization for Nuclear Research (CERN), Geneva, Switzerland\\
% ...
$ ^{38}$Nikhef National Institute for Subatomic Physics, Amsterdam, The Netherlands\\
% ...
$ ^{43}$H.H. Wills Physics Laboratory, University of Bristol, Bristol, United Kingdom\\
% ...
$ ^{49}$Oliver Lodge Laboratory, University of Liverpool, Liverpool, United Kingdom\\
% ...
$ ^{n}$LIFAELS, La Salle, Universitat Ramon Llull, Barcelona, Spain\\
% ...
}
\end{flushleft}
%%%%%%%%%%%%%%%%%%%%%%%%%%%%%%%%%%%%%%%%%%
}
\begin{abstract}
\noindent 
A search for  \Bsmumu and \Bdmumu decays is performed 
using $1.0$\invfb of $pp$ collision data collected at $\sqrt{s}=7$\,TeV
with the LHCb experiment at the Large Hadron Collider. 
For both decays the number of observed events is consistent with expectation 
from background and Standard Model signal predictions. 
Upper limits on the branching fractions are determined to 
be \BRof \Bsmumu$< 4.5 \,(3.8)  \times 10^{-9}$  and \BRof\Bdmumu $<1.0\, (0.81) \times 10^{-9}$ at 95\,\% (90\,\%) confidence level.

%\begin{center}
%\emph{To be submitted to Physical Review Letters}
%\end{center}

\end{abstract}


\vspace*{-1cm}
\hspace{-9cm}
\mbox{\Large EUROPEAN ORGANIZATION FOR NUCLEAR RESEARCH (CERN)}

\vspace*{0.1cm}
\hspace*{-9cm}
\begin{tabular*}{16cm}{lc@{\extracolsep{\fill}}r}
\ifthenelse{\boolean{pdflatex}}% Logo format choice
{\vspace*{-3.2cm}\mbox{\!\!\!\includegraphics[width=.14\textwidth]{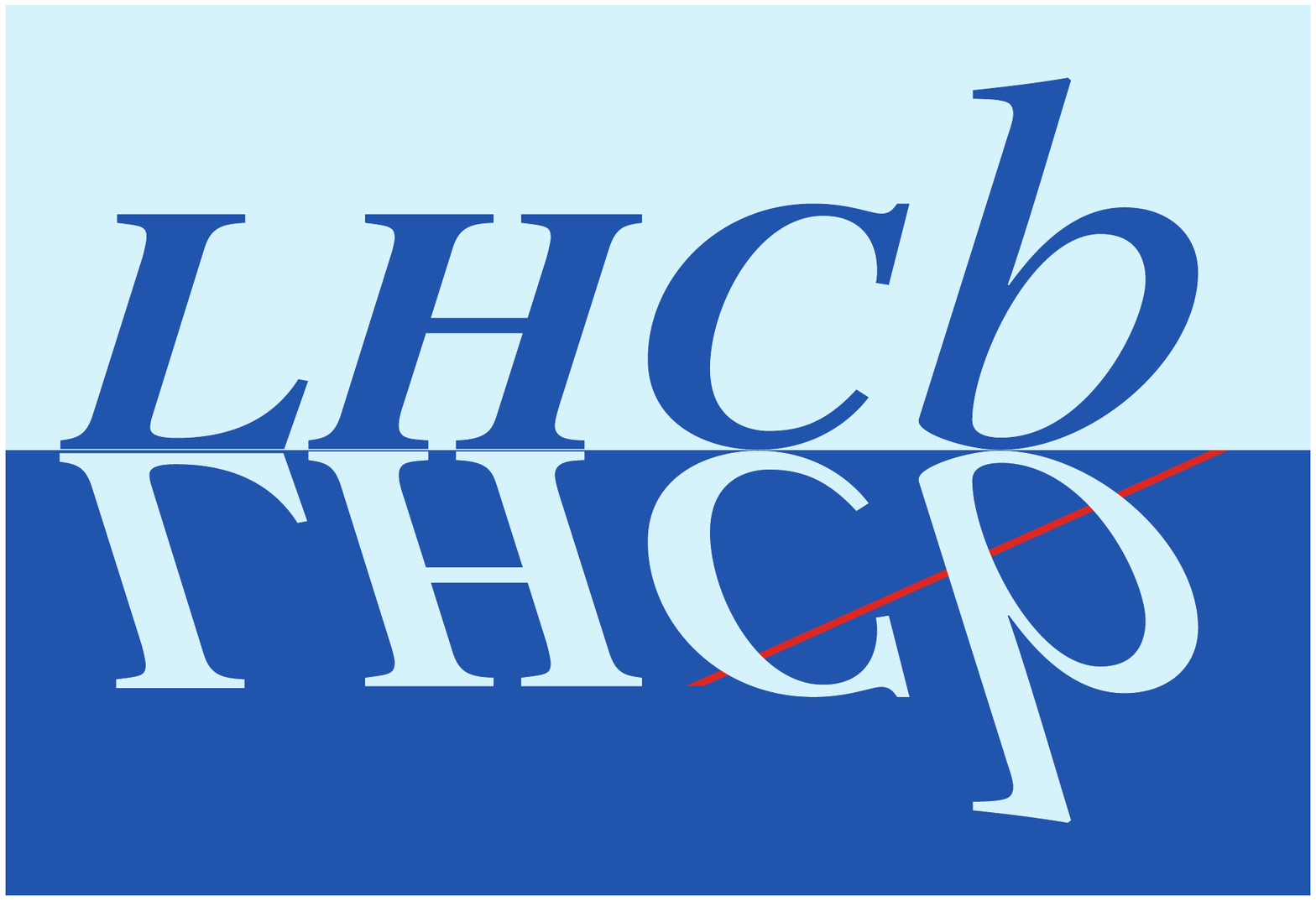}} & &}%
{\vspace*{-1.2cm}\mbox{\!\!\!\includegraphics[width=.12\textwidth]{lhcb-logo.eps}} & &}%
\\
 & & LHCb-PAPER-2012-007 \\
 & & CERN-PH-EP-2012-072 \\ % ID 
 & & \today \\ % Date - Can also hardwire e.g.: 23 March 2010
%  & & \\
% not in paper \hline
\end{tabular*}
\vspace*{1cm}

\maketitle

%\nofootinbib

%\input{introduction}

%----------------
% Introduction
%----------------

\noindent
Flavor changing neutral current
(FCNC) processes are highly suppressed in the Standard Model (SM) and thus 
constitute a stringent test of the current description of particle physics. 
Precise predictions of the branching fractions of 
the FCNC decays \Bsmumu and \Bdmumu, \BRof \Bsmumu = $(3.2 \pm
0.2) \times 10^{-9}$ and \BRof \Bdmumu = $(0.10 \pm 0.01) 
\times 10^{-9}$~\cite{Buras2010mh,Buras:2010wr} 
make these modes powerful 
probes in the search for deviations from the
SM, as contributions from new processes or new 
heavy particles can significantly modify these values.
Previous searches~\cite{d0_PLB,cdf_prl,cms,lhcbpaper2}
already constrain possible deviations from the SM predictions, with the lowest published limits from the LHCb collaboration:
\mbox{\BRof \Bsmumu $< 1.4 \times 10^{-8}$} and \mbox{\BRof \Bdmumu $< 3.2 \times 10^{-9}$} at 95\% Confidence Level (\CL).

In this Letter, we report an analysis of the $pp$ collision 
data recorded in  2011 by the LHCb experiment 
corresponding to an integrated luminosity of  1.0\invfb.
This dataset includes the 0.37\invfb used in the previous 
analysis~\cite{lhcbpaper2}.
In addition to the larger dataset, improvements include an updated
event selection, an optimized binning in the discriminating variables, 
and a reduction of the peaking background.
The data already analyzed in Ref.~\cite{lhcbpaper2} were reprocessed and,
to avoid any potential bias, all the events in the signal region were blinded
until all the analysis choices were finalized.

%--------------------------------
% The LHCb detector and trigger
%--------------------------------
The \lhcb detector~\cite{LHCbdetector} is a single-arm forward
spectrometer covering the pseudo-rapidity range \mbox{$2<\eta<5$}. The
detector includes a high precision tracking system consisting of a
silicon-strip vertex detector,
a large-area silicon-strip detector located upstream of a dipole
magnet with a bending power of about $4{\rm\,Tm}$, and three stations
of silicon-strip detectors and straw drift tubes placed
downstream. The combined tracking system has a momentum resolution
$\Delta p/p$ that varies from 0.4\,\% at 5\gevc to 0.6\,\% at 100\gevc. 
Two ring-imaging Cherenkov detectors (RICH) are used to identify charged 
particles. Photon, electron and hadron
candidates are identified by a calorimeter system consisting of
scintillating-pad and pre-shower detectors, an electromagnetic
calorimeter and a hadronic calorimeter. Muons are identified by 
alternating layers of iron and multiwire proportional chambers.

The trigger consists of a hardware stage, based
on information from the calorimeter and muon systems, followed by a
software stage (HLT) that applies a full event reconstruction.
Events with muon final states are 
triggered using two hardware trigger decisions: the single-muon decision 
(one muon candidate with transverse momentum 
$p_{\rm T}> 1.5$\gevc), and the di-muon 
decision (two muon candidates with
$p_{{\rm T},1}$ and $p_{{\rm T},2}$ such that 
$\sqrt{p_{{\rm T},1}p_{{\rm T},2}} > 1.3$\gevc). 
All tracks in the HLT are required to have a $p_{\rm T}> 0.5$\gevc.
The single muon trigger decision in the HLT
selects tracks with an impact parameter \IP$>0.1$\,mm 
and $p_{\rm T}>1.0$\gevc.
The di-muon trigger decision requires $\mu^+\mu^-$ pairs 
with an invariant mass $m_{\mu\mu} > 4700$\mevcc.   
Another trigger decision, designed to select $J/\psi$ mesons, requires 
$2970 < m_{\mu\mu} < 3210$\mevcc. 
Events with purely hadronic final states are triggered by the hardware trigger 
if there is a calorimeter cluster with transverse energy $E_{\rm T} > 
3.5$\,GeV.
HLT trigger decisions selecting generic $b$-hadrons decays provide high 
efficiency for such final states.

%---------------
% selection
%---------------
The \Bmumu selection requires two high quality muon candidates 
displaced with respect to any primary $pp$ interation point (primary vertex, 
PV). 
The di-muon secondary vertex (SV) is required to be well measured (with a $\chi^2$ per degree of freedom smaller than 9.0), 
downstream, and separated from the
PV by a distance-of-flight significance greater than 15.
When more than one PV is
reconstructed, the one giving the minimum IP significance for the $B$ candidate is chosen. 
Only candidates with ${\rm IP}/\sigma({\rm IP})<5$ are kept.
Combinations with poorly reconstructed tracks are removed
by requiring \mbox{$p<$ 500\,GeV/$c$} and $0.25<p_{\rm T}<40$\gevc for 
all tracks from the selected candidates.
Only \B candidates with decay times smaller than $9 \times \tau(\Bs)$~\cite{PDG} are kept.
Finally, according to the simulation, approximately 90\% of di-muon candidates coming from elastic di-photon production are removed by requiring a minimum $p_{\rm T}$ of the $B$ candidate of 500\mevc. The surviving background mainly comprises random combinations of muons from semileptonic b-hadrons decays (\bbdim, where $X$ is any other set of particles).

Three channels, \BuJpsiK, \BsJpsiPhi, and $B^0 \to K^+ \pi^-$ 
(inclusion of charged conjugated processes is implied throughout this Letter) 
serve as normalization modes. 
The first two have trigger 
and muon identification efficiencies similar to those of the signal, but 
a different number of tracks in the final state. The third channel has a 
similar 
topology, but is selected by different triggers. The selection of 
these channels is designed to be as similar as possible to that of 
the signal to reduce the impact of common systematic uncertainties.
An inclusive \Bhh sample (where $h, h'$ can be a  
pion or a kaon) is the main control sample. 
The selection is the same as for \Bmumu signal candidates, 
except for the muon identification 
requirement. 
To ensure similar selection efficiencies for the \BdKpi and \Bmumu channels, tracks from the \BdKpi decay are required to be in the muon detector acceptance.
The $J/\psi \to \mu^+ \mu^-$ decay in the \BuJpsiK and \BsJpsiPhi normalization 
channels is also selected as \Bmumu signal, 
except for the requirements on its IP and mass.
Kaon candidates are required to be identified by the RICH detectors and 
to pass IP selection criteria.

A multivariate selection (MVS), based on a boosted decision tree~\cite{tmva}, removes 80\,\% of the residual background, while retaining 92\,\% of the signal. Applying this selection improves the performance of the main
multivariate algorithm described below. The six variables entering the MVS, ordered by their background rejection power, are:  
the angle between the direction of the momentum of the \B candidate and the direction 
defined by the vector joining the secondary and the primary vertices, 
the \B candidate IP and its vertex $\chi^2$, the minimum IP of the muons with respect to any PV, 
the minimum distance between the two daughter tracks and the $\chi^2$ of the SV.  
The \Bhh mass sidebands have been used to check that the distribution of the MVS output is similar for data and simulation.
The same selection is applied (using, when necessary, slightly modified variable
definitions) to the normalization samples. 
The efficiencies for the signal and the normalization samples are equal within 0.2\,\% according to the simulation.

In total, 17\,321 muon pairs with invariant mass between 4900 and 6000\mevcc 
pass the trigger and selection requirements. 
Given the measured $b \bar{b}$ 
cross-section~\cite{bbxsection} and assuming SM rates, 
this data sample is expected to contain 
$11.6$ $B^0_{s}\to \mu^+\mu^-$ and $1.3$ $B^0 \to \mu^+\mu^-$ decays.

The selected candidates are classified in a binned two-dimensional space
formed by the di-muon invariant mass and the output of another boosted decision tree, 
described in detail below. In the following we employ ``BDT'' to indicate the algorithm or its ouput, 
depending on the context.

The invariant mass line shape of the signal events 
is described by a Crystal Ball function~\cite{crystalball}.
The peak values for the \Bs and \Bd mesons, \mBs and \mBd, 
are obtained 
from  the \BsKK and \BdKpi samples~\cite{LHCb_paper}.
The resolutions are extracted from data with  
a power-law interpolation between the measured resolutions of charmonium and 
bottomonium 
resonances 
decaying into two muons.
Each resonance is fitted with the sum of two Crystal Ball functions with 
common mean
values and resolutions, but different parameters describing the tails. 
The results of the interpolation at \mBs and \mBd are
$ \sigma(\mBs)  =  24.8 \pm 0.8 \mevcc$ and
$\sigma(\mBd)  =  24.3 \pm 0.7 \mevcc$. 
They are in agreement with those found using \BdKpi and \BsKK exclusive decays.
The transition point of the radiative tail is obtained from simulated \Bsmumu events re-weighted to reproduce the mass resolution measured in data.

%--------------------------------
% Multivariate analysis
%--------------------------------
Geometrical and kinematic information not fully exploited in the selection 
is combined via the BDT
for which nine variables are employed~\cite{lhcbpaper2}. Ordered by their background rejection power, they are: 
the \B candidate IP, the minimum IP significance, the
sum of the degrees of isolation of the muons 
(the number of good two-track vertices a muon can make with other tracks 
in the event), the \B candidate decay time, $p_{\rm T}$, 
and degree of isolation~\cite{cdf_iso},
the distance of closest approach between the two muons, 
the minimum $p_{\rm T}$ of the muons,
and the cosine of the angle between the muon momentum in the di-muon 
rest frame and the vector perpendicular to the \B candidate 
momentum and to the beam axis. 
No data were used for the choice of the variables and the subsequent 
training of the BDT, to avoid biasing the results. 
Instead the BDT was trained using simulated samples 
(\Bmumu for signal and \bbdim for background). 
The BDT output is independent of the invariant mass for signal 
inside the search window. 
It is defined such that for the signal it is approximately uniformly 
distributed  between zero and one, while for the background it peaks at zero.

The probability for a signal event to have a given BDT value is 
obtained from data using an inclusive \Bhh sample. Only events triggered independently of the 
presence of any track from the signal candidates are considered. 
The number of \Bhh signal events in each BDT bin is determined by 
fitting the $hh'$ invariant mass distribution.
The maximum spread in the fractions of the yields going into each bin, obtained by fitting the same dataset with different signal and background models, 
is used  to evaluate the systematic
uncertainty on the signal BDT probability distribution 
function~\cite{lhcbpaper2}.

The binning of the BDT and invariant mass distributions is re-optimized with respect to Ref.~\cite{lhcbpaper2}, 
using simulation, to maximize the separation between the median of the test statistic distribution 
expected for background and SM \Bsmumu signal, and that expected for background only. 
The chosen number and size of the bins are a compromise between maximizing the number of bins and the 
necessity to have enough \Bhh events 
to calibrate the  \Bsmumu BDT and enough background in the mass sidebands (see below) in each bin to estimate 
the combinatorial background in the \Bs and \Bd mass regions.
The BDT range is thus divided into eight bins (see Table~\ref{tab:data_bsdmm}) and 
the invariant mass 
range into nine bins
with boundaries are defined by $m_{B^{0}_{(s)}} \pm 18, 30, 36, 48, 60$\mevcc.
This binning improves the test statistic separation by about 14\,\% at the SM rate with respect to Ref.~\cite{lhcbpaper2}; over 97\,\% of this separation comes from the bins with BDT$>0.5$.

We select events in the invariant mass range \mbox{$4900\mevcc, 6000\mevcc]$}. The boundaries of the signal regions are defined as 
$m_{B^{0}_{(s)}} \pm 60$\mevcc.
The low-mass sideband is potentially polluted by  
cascading $b \to c \mu \nu \to \mu \mu X$ 
decays below 4900\mevcc and peaking background from \Bhh candidates with the 
two hadrons misidentified 
as muons above 5000\mevcc. 
%The BDT and invariant mass shapes for the combinatorial background inside the 
%signal regions are determined from data by interpolating the number of  
%expected events with a fit in each BDT bin to an exponential function to 
%the events in the low-mass range [4900, 5000]\mevcc and in the full 
%high-mass sidebands.
%--
The number of expected combinatorial background events in each BDT and invariant mass bin 
inside the signal regions
is determined from data by fitting to an exponential function events in 
the mass sidebands defined by [4900, 5000] \mevcc and [\mBs+60 \mevcc, 6000 \mevcc].
The systematic uncertainty on the estimated number of combinatorial 
background events is computed by fluctuating 
with a Poissonian distribution the number of events measured in the sidebands, 
and by varying within $\pm 1 \sigma$ the value of the exponent.
%--
As a cross-check, another model, 
the sum of two exponential functions,
has been used to fit the events in different ranges of sidebands providing 
consistent background estimates inside the signal regions. 
An additional systematic uncertainty is introduced where the yields in the signal 
regions differ by more than $1\,\sigma$ between the fit models.

Peaking backgrounds from \Bhh events have been evaluated by folding the $K \to \mu$ and 
$\pi \to \mu$ misidentification rates extracted from a $D^0 \to  K^- \pi^+$ sample from data in bins of $p$ and $p_T$ into the 
spectrum of selected simulated \Bhh events.
The mass line shape of the peaking background is obtained from a simulated 
sample of doubly-misidentified \Bhh events. 
In total, $ 0.5^{+0.2}_{-0.1}$ ($2.6^{+1.1}_{-0.4}$) doubly-misidentified \Bhh events are expected in the \Bs (\Bd) signal mass windows. 
The contributions of \mbox{$B^+_c \to J/\psi(\mu^+ \mu^-) \mu^+ \nu$} and $B^0_s \to \mu^+ \mu^- \gamma$ exclusive decays
have been found to be negligible with respect to the combinatorial and \Bhh backgrounds.

%------------------------
% Normalization
%------------------------
The \Bsmumu and \Bdmumu yields are translated 
into branching fractions using
\begin{eqnarray}
{\cal B} &=&  {\cal B}_{\rm norm} \times
\frac{\rm \epsilon_{norm}}{\rm \epsilon_{sig}}  \times
\frac{ f_{\rm norm}}{ f_{d(s)}} \times 
\frac{N_{\Bmumu}}{N_{\rm norm}}  \nonumber \\
& = & \alpha^{\rm norm}_{\Bmumu} \times N_{\Bmumu},
\label{eq:normalization}
\end{eqnarray}
where $f_{d(s)}$ and $f_{\rm norm}$ are the probabilities
that a $b$ quark fragments into a $B^0_{(s)}$ and into the hadron involved
in the given normalization mode respectively. 
We use $f_s/f_d = 0.267^{+0.021}_{-0.020}$~\cite{LHCb-PAPER-2011-018} and we assume $f_d=f_u$. 
With ${\cal B}_{\rm norm}$ we indicate the branching fraction 
and with $N_{\rm norm}$ the number of signal events in the normalization 
channel obtained from a fit to the invariant mass distribution.
The efficiency ${\rm \epsilon_{sig(norm)}}$ for the signal (normalization channel) is
the product of 
the reconstruction efficiency of all the final state particles of the decay 
including the geometric acceptance of the detector, 
the selection efficiency for reconstructed events, and
the trigger efficiency for reconstructed and selected events. 
%--
The ratio of acceptance and reconstruction efficiencies are computed using the Monte Carlo
simulation. The differences between the simulation and data are included as systematic uncertainties. 
The selection efficiencies are determined using Monte Carlo simulation and cross-checked with data.
Reweighting techniques have been used for all the Monte Carlo distributions that  do not match 
those from data. The trigger efficiency is evaluated with data driven techniques.
%--
Finally, $N_{\Bmumu}$ is the number of observed 
signal events.
The observed numbers of \BuJpsiK, \BsJpsiPhi and \BdKpi candidates are
$340\,100\pm4500$, $19\,040\pm160$ and $10\,120\pm920$, respectively.
The three normalization factors are in agreement within the uncertainties and their
weighted average, taking correlations into account,
gives
$\alpha^{\rm norm}_{\Bsmumu}= (3.19 \pm 0.28) \times 10^{-10}$ and
$\alpha^{\rm norm}_{\Bdmumu}= (8.38 \pm 0.39) \times 10^{-11}$.

For each bin in the two-dimensional space formed by the invariant mass and the 
BDT we count the number of candidates observed in the 
data, and compute the expected number of signal and background events.

%--
{The systematic uncertainties in the background and signal predictions in each bin 
are computed by fluctuating the mass and BDT shapes and the normalization factors
along the Gaussian  distributions defined by their associated  uncertainties.
The inclusion of the systematic uncertainties increases the \Bdmm and \Bsmm upper limits by less than $\sim 5\%$.}
%--

\begin{figure}[htb]
  \begin{center}
    \includegraphics*[width=0.23\textwidth]{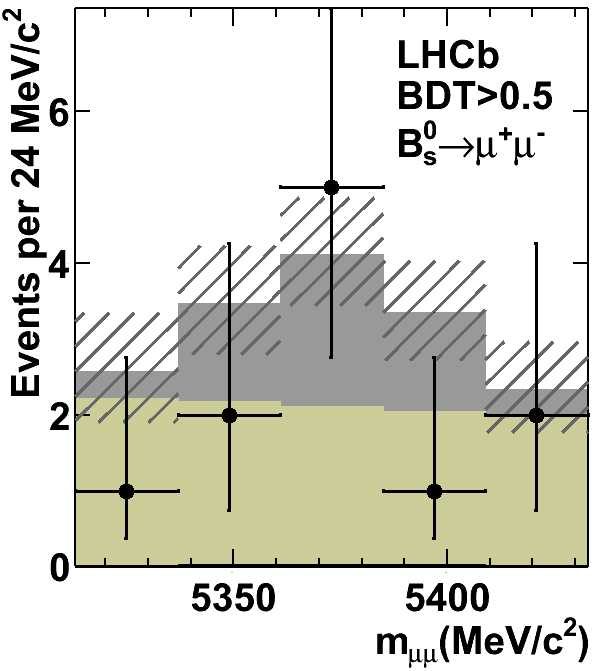}
    \includegraphics*[width=0.23\textwidth]{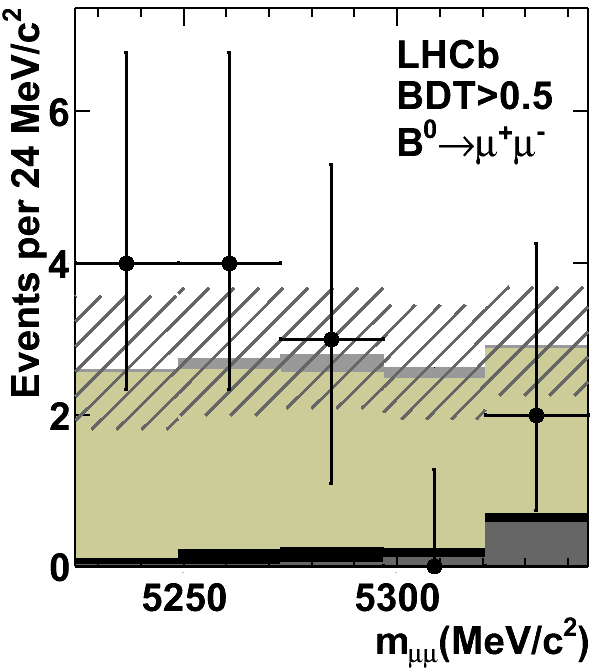}
  \end{center}
\caption{Distribution of selected candidates (black points) in the (left) \Bsmumu  and (right) \Bdmumu 
mass window for BDT$>$0.5, and expectations for, from the top, \Bmumu SM signal (gray), combinatorial background (light gray), \Bhh background (black), and cross-feed of the two modes (dark gray). The hatched area depicts the uncertainty on the sum of the expected contributions.}
\label{fig:fondo_bsd}
\end{figure}

The results for \Bsmm and \Bdmumu decays, integrated over all mass bins in the 
corresponding signal region, 
are summarized in Table~\ref{tab:data_bsdmm}. The distribution of the 
invariant mass for BDT$>$0.5 is shown in Fig.~\ref{fig:fondo_bsd} for 
\Bsmumu and \Bdmumu candidates. 

\begin{center}
\begin{table*}[htb]
\centering
\caption[]{Expected combinatorial background,  \Bhh background, cross-feed, and signal events assuming SM predictions, together with the number of observed events in the \Bsmumu and \Bdmumu mass signal regions, in bins of BDT.} 
    \label{tab:data_bsdmm}
\begin{tabular}{cccccccccc}
\hline\hline
Mode & BDT bin & 0.0 -- 0.25 & 0.25 -- 0.4 & 0.4 -- 0.5 & 0.5 -- 0.6 & 0.6 -- 0.7 & 0.7 -- 0.8 & 0.8 -- 0.9 & 0.9 -- 1.0 \TTstrut\BBstrut \\
\hline
\Bsmumu &  Exp. comb. bkg & $1889^{+38}_{-39}$ & $57^{+11}_{-11}$ & $15.3^{+3.8}_{-3.8}$ & $4.3^{+1.0}_{-1.0}$ & $3.30^{+0.92}_{-0.85}$ & $1.06^{+0.51}_{-0.46}$ & $1.27^{+0.53}_{-0.52}$ & $0.44^{+0.41}_{-0.24}$  \TTstrut\\
 &  Exp. peak. bkg & $0.124^{+0.066}_{-0.049}$ & $0.063^{+0.024}_{-0.018}$ & $0.049^{+0.016}_{-0.012}$ & $0.045^{+0.016}_{-0.012}$ & $0.050^{+0.018}_{-0.013}$ & $0.047^{+0.017}_{-0.013}$ & $0.049^{+0.017}_{-0.013}$ & $0.047^{+0.018}_{-0.014}$  \TTstrut\\
 &  Exp. signal & $2.55^{+0.70}_{-0.74}$ & $1.22^{+0.20}_{-0.19}$ & $0.97^{+0.14}_{-0.13}$ & $0.861^{+0.102}_{-0.088}$ & $1.00^{+0.12}_{-0.10}$ & $1.034^{+0.109}_{-0.095}$ & $1.18^{+0.13}_{-0.11}$ & $1.23^{+0.21}_{-0.21}$  \TTstrut\\
 &  Observed & $1818$ & $39$ & $12$ & $6$ & $1$ & $2$ & $1$ & $1$  \TTstrut\BBstrut\\
\hline
\Bdmumu &  Exp. comb. bkg & $2003^{+42}_{-43}$ & $61^{+12}_{-11}$ & $16.6^{+4.3}_{-4.1}$ & $4.7^{+1.3}_{-1.2}$ & $3.52^{+1.13}_{-0.97}$ & $1.11^{+0.71}_{-0.50}$ & $1.62^{+0.76}_{-0.59}$ & $0.54^{+0.53}_{-0.29}$  \TTstrut\\
   & Exp. peak. bkg & $0.71^{+0.36}_{-0.26}$ & $0.355^{+0.146}_{-0.088}$ & $0.279^{+0.110}_{-0.068}$ & $0.249^{+0.099}_{-0.055}$ & $0.280^{+0.109}_{-0.062}$ & $0.264^{+0.103}_{-0.057}$ & $0.275^{+0.108}_{-0.060}$ & $0.267^{+0.106}_{-0.069}$  \TTstrut\\
  & Exp. cross-feed & $0.40^{+0.11}_{-0.12}$ & $0.193^{+0.033}_{-0.030}$ & $0.153^{+0.023}_{-0.021}$ & $0.136^{+0.017}_{-0.015}$ & $0.158^{+0.019}_{-0.017}$ & $0.164^{+0.019}_{-0.017}$ & $0.187^{+0.022}_{-0.020}$ & $0.194^{+0.036}_{-0.033}$  \TTstrut\\
  &  Exp. signal & $0.300^{+0.086}_{-0.090}$ & $0.145^{+0.027}_{-0.024}$ & $0.115^{+0.020}_{-0.017}$ & $0.102^{+0.014}_{-0.013}$ & $0.119^{+0.017}_{-0.015}$ & $0.123^{+0.016}_{-0.015}$ & $0.140^{+0.019}_{-0.017}$ & $0.145^{+0.030}_{-0.026}$  \TTstrut\\
  & Observed & $1904$ & $50$ & $20$ & $5$ & $2$ & $1$ & $4$ & $1$  \TTstrut\BBstrut\\ \hline\hline

\end{tabular}
\end{table*}
  \end{center}

The compatibility of the observed distribution of events  
with that expected for a given branching fraction 
hypothesis is computed using the \CLs method~\cite{Read_02}.
The method provides \CLsb, a measure of the 
compatibility of the observed distribution with the signal plus background 
hypothesis, \CLb, a measure of the compatibility with the background-only 
hypothesis, and \mbox{$\CLs=\CLsb/\CLb$}.

\begin{figure*}[!htb]
\centering
\includegraphics[width=0.45\textwidth]{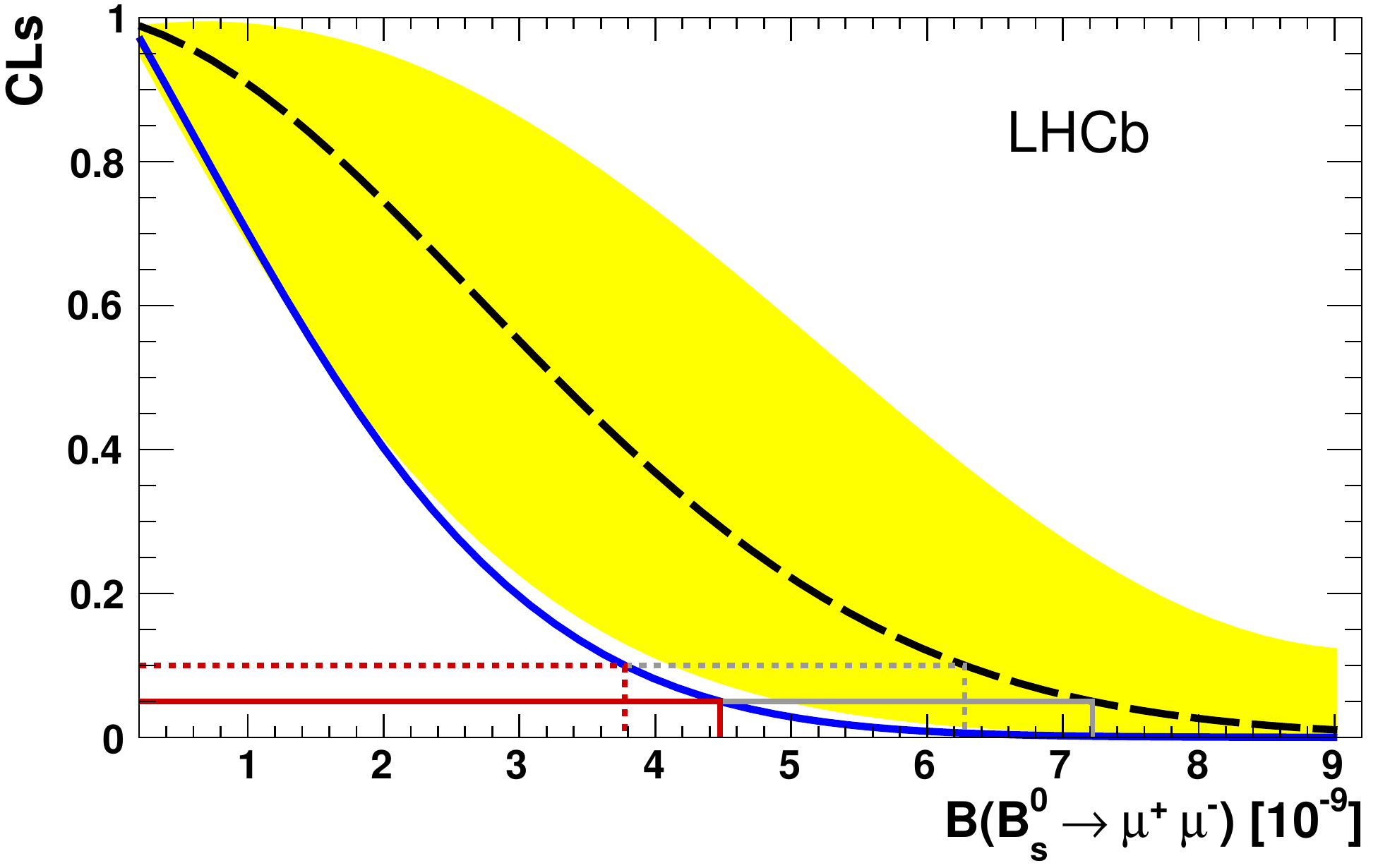}
\includegraphics[width=0.45\textwidth]{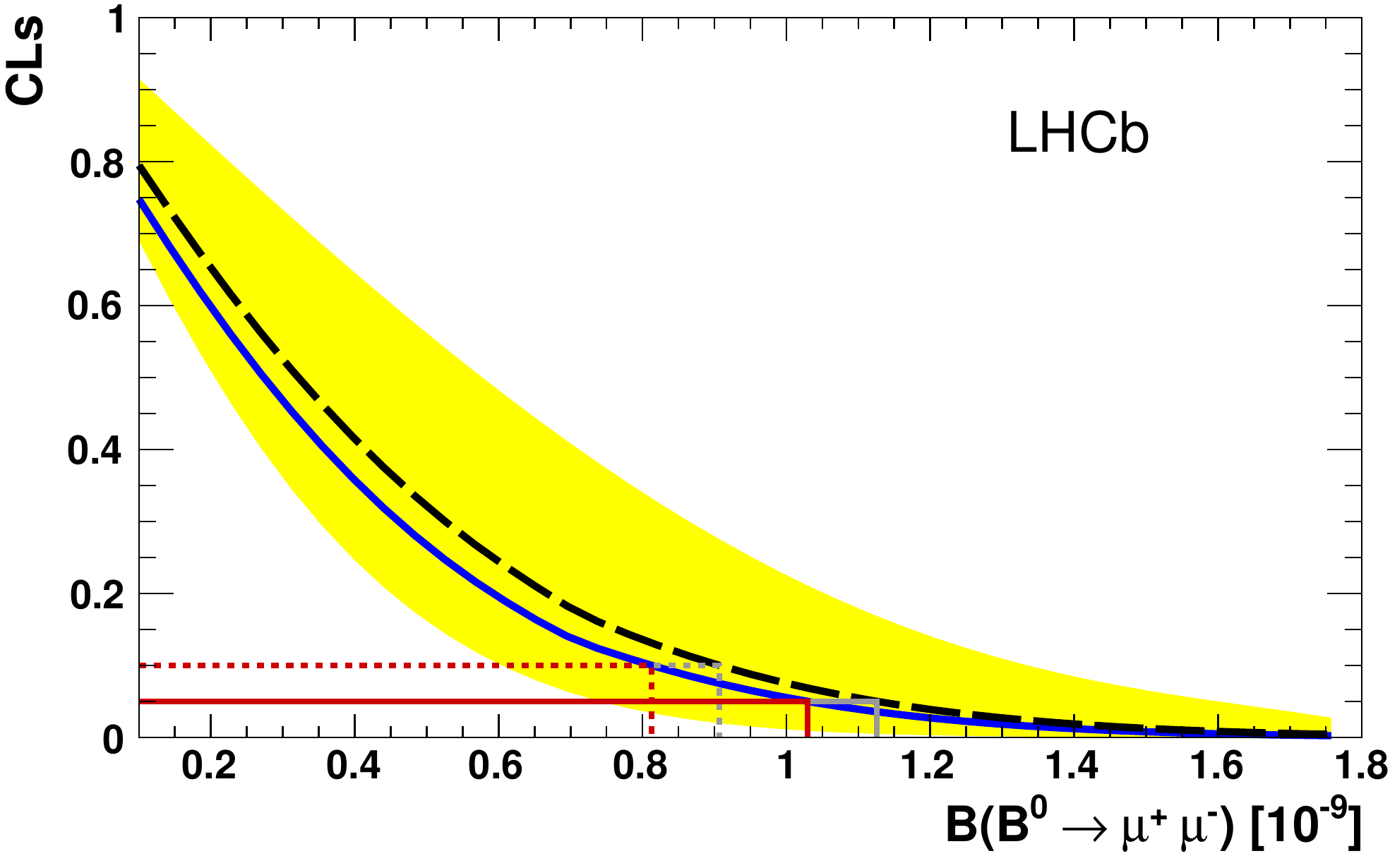}
\vspace{-4mm}
\caption
{
 \CLs as a function of the assumed \BF\ for (left) \Bsmumu and (right) 
\Bdmumu decays.
The long dashed black curves are the medians of the expected \CLs\ distributions for \Bsmumu, if background and SM signal were observed, and for \Bdmumu, if background only was observed.
The yellow  areas cover, for each \BF, 34\% of the expected \CLs distribution on each side of its median.
The solid blue curves are the observed \CLs. The upper limits at 90\,\% (95\,\%) \CL are indicated by the dotted (solid) horizontal lines in red (dark gray) for the observation and in gray for the expectation.}
\label{fig:cls_bsbd}
\end{figure*}

The expected and observed \CLs values are shown in Fig.~\ref{fig:cls_bsbd} for the \Bsmumu and \Bdmumu channels,
each as a function of the assumed branching fraction.
The expected and measured limits for \Bsmumu and \Bdmumu at 90\,\% 
and 95\,\% \CL are shown in Table~\ref{tab:bds_results}.
The expected limits are computed allowing the 
presence of \Bmumu events according to the SM branching fractions, including 
cross-feed between the two modes. 

The comparison of the distributions of observed events and expected
background events results in a p-value \mbox{$(1-\CLb)$} of 18\,\% (60\,\%)
for the \Bsmumu \mbox{(\Bdmumu)} decay, where the \CLb values are those
corresponding to $\CLsb=0.5$.

\begin{table}[!htb]
\caption{Expected and observed limits on the \Bmumu branching fractions. 
}
\label{tab:bds_results}
\begin{center}
\begin{tabular}{@{}llcc@{}}
\hline\hline 
         Mode & Limit & at 90\,\% \CL & at 95\,\% \CL\TTstrut\BBstrut\\ 
\hline 
\Bsmumu      & Exp. bkg+SM           &  $6.3 \times 10^{-9} $  & $ 7.2  \times 10^{-9} $\TTstrut\\ 
             & Exp. bkg              &  $2.8 \times 10^{-9} $  & $ 3.4  \times 10^{-9} $   \\ 
             & Observed              &  $3.8 \times 10^{-9} $  & $ 4.5  \times 10^{-9} $   \\ 
\hline
\Bdmumu      & Exp. bkg              &  $0.91 \times 10^{-9}$  & $ 1.1 \times 10^{-9}$\TTstrut\\ 
             & Observed              &  $0.81 \times 10^{-9}$  & $ 1.0 \times 10^{-9}$   \\ 
\hline \hline
\end{tabular}
\end{center}
\end{table}

A simultaneous unbinned likelihood fit to the mass projections in the eight BDT bins has been performed to determine 
the  \Bsmumu
branching fraction. The signal fractional yields in BDT bins are
constrained to the BDT fractions calibrated with the \Bhh sample.
The fit gives \BRof \Bsmumu = $(0.8^{+1.8}_{-1.3}) \times 10^{-9}$,
where the central value is extracted from the maximum of the logarithm of the
profile likelihood and the uncertainty reflects the interval corresponding to a
change of 0.5. Taking the result of the fit as a posterior, with a positive branching fraction as a flat prior, the probability
for a measured
value to fall between zero and the SM expectation is 82\,\%, according to the simulation.
The one-sided 90\,\%, 95\,\% \CL limits, and the compatibility with the SM
predictions obtained from the likelihood, are in agreement with the \CLs
results. 
The results of a fully unbinned likelihood fit method are in agreement
within uncorrelated systematic uncertainties. The largest systematic
uncertainty is due to the parametrization of the combinatorial background
BDT.

In summary, a search for the rare decays \Bsmumu
and \Bdmumu has been performed on a data sample corresponding to an
integrated luminosity of 1.0\invfb.
These results supersede those of our previous publication~\cite{lhcbpaper2}  and are statistically independent of those obtained from data collected in 2010~\cite{LHCb_paper}.
The data 
are consistent with both the background-only hypothesis and the combined background plus SM signal 
expectation at the 1\,$\sigma$ level.
For these modes we  set the most stringent upper limits to date: \BRof \Bsmumu $< 4.5 \times 10^{-9}$ and 
\BRof \Bdmumu $< 1.03 \times 10^{-9}$ at 95\,\% \CL.

%\input{acknowledgments}
%\clearpage

We express our gratitude to our colleagues in the CERN accelerator
departments for the excellent performance of the LHC. We thank the
technical and administrative staff at CERN and at the LHCb institutes,
and acknowledge support from the National Agencies: CAPES, CNPq,
FAPERJ and FINEP (Brazil); CERN; NSFC (China); CNRS/IN2P3 (France);
BMBF, DFG, HGF and MPG (Germany); SFI (Ireland); INFN (Italy); FOM and
NWO (The Netherlands); SCSR (Poland); ANCS (Romania); MinES of Russia and
Rosatom (Russia); MICINN, XuntaGal and GENCAT (Spain); SNSF and SER
(Switzerland); NAS Ukraine (Ukraine); STFC (United Kingdom); NSF
(USA). We also acknowledge the support received from the ERC under FP7
and the Region Auvergne.

Note added in proof: while this paper was in preparation, the CMS 
collaboration released the results of an updated search for these 
channels~\cite{Chatrchyan:2012rg}.

\bibliographystyle{LHCb}
\bibliography{bsmumureferences}

\end{document}